# Enhancing spin-orbit torque by strong interfacial scattering from ultra-thin insertion layers

Lijun Zhu,[1] Lujun Zhu,[2] Shengjie Shi,[1] Manling Sui,[3] D. C. Ralph,[1,4] R. A. Buhrman[1]
1. Cornell University, Ithaca, NY 14850
2. College of Physics and Information Technology, Shaanxi Normal University, Xi'an, 710062, China
3. Institute of Microstructure and Property of Advanced Materials, Beijing University of Technology, Beijing 100124, China
4. Kavli Institute at Cornell, Ithaca, New York 14853, USA

Increasing dampinglike spin-orbit torque (SOT) is both of fundamental importance for enabling new research into spintronics phenomena and also technologically urgent for advancing low-power spin-torque memory, logic, and oscillator devices. Here, we demonstrate that enhancing interfacial scattering by inserting ultra-thin layers within a spin Hall metals with intrinsic or side-jump mechanisms can significantly enhance the spin Hall ratio. The dampinglike SOT was enhanced by a factor of 2 via sub-monolayer Hf insertion, as evidenced by both harmonic response measurements and current-induced switching of in-plane magnetized magnetic memory devices with the record low critical switching current of ~73 μA (switching current density ≈ 3.6×10$^6$ A/cm$^2$). This work demonstrates a very effective strategy for maximizing dampinglike SOT for low-power spin-torque devices.

## 1. Introduction

Spin-orbit torques (SOTs) generated by the spin Hall effect (SHE) can efficiently switch thin-film nanomagnet devices [1-4], excite magnetization oscillations [5], and drive skyrmion and chiral domain wall displacement [7,8]. Increasing SOT efficiencies is of great importance for enabling new research into spintronics phenomena [1-9] and for advancing technological applications of SOTs [10-13]. Of particular interest in this effort is to develop heavy metals (HMs) that can simultaneously provide a large damping-like SOT efficiency per current density ($\xi_{DL}^j$), easy growth, good chemical/thermal stability, and the capability to be readily integrated into complex experimental configurations and/or into manufacturing processes. A good representative of such HMs is Pt, which has giant spin Hall conductivity ($\sigma_{SH}$) arising from the Berry curvature of its band structure [14,15]. For the SHE, $\xi_{DL}^j \equiv (2e/\hbar)T_{int}\sigma_{SH}\rho_{xx}$ with $e$, $\hbar$, $\rho_{xx}$, and $T_{int}$ being the elementary charge, the reduced Planck constant, the HM resistivity, the spin transparency of the HM/FM interface [9]. $\xi_{DL}^j$ for Pt/ferromagnet (FM) systems is ~0.08 where $\rho_{xx}$ = 20 μΩ cm [16]. Recently, impurity scattering has been demonstrated to increase $\xi_{DL}^j$ via enhancing $\rho_{xx}$ [17-20]. However, in all the previous work the increase of $\xi_{DL}^j$ was limited (e.g., to $\xi_{DL}^j$ = 0.12-0.3 for 4 nm Pt alloys) due to a fast decrease in $\sigma_{SH}$ with doping level [19] or/and only a weak enhancement of $\rho_{xx}$ [17,18]. Exploring new enhancement strategies that can better optimize the trade-offs between $\rho_{xx}$ and $T_{int}\sigma_{SH}$ is of both fundamental interest and technological urgency (e.g., for low-power magnetic memories, logic, and oscillators).

In this work, we report that introducing strong interfacial electron scattering via the insertion of sub-monolayers of Hf into Pt can enhance $\rho_{xx}$ of a ~4 nm Pt layer by a factor of 5, which beneficially results in 100% enhancement of $\xi_{DL}^j$ (up to 0.37). The increase in $\xi_{DL}^j$ by the ultrathin insertion layers is approximately twice as effective as a uniform alloying of Hf into Pt. This giant enhancement of $\xi_{DL}^j$ by Hf insertion layers is reaffirmed by the deterministic switching of in-plane magnetic tunnel junctions (MTJs) at a low critical current of ≈ 73 μA (current density ≈ 3.6×10$^6$ A/cm$^2$) in absence of the assistance of thermal fluctuations.

## 2. Results and discussions
### 2.1 Enhancing resistivity by interfacial scattering

The main idea of this work is schematically shown in Fig. 1. In a single metallic layer of Pt, that is not too thin, e.g. 4 nm as typically used for spin-torque magnetic random access memories (MRAMs)[10], the resistivity arises mainly from the electron scattering by impurities and thermal phonons inside the Pt layer and is hence relatively low, e.g. 20-50 μΩ cm at room temperature [16-21]. In contrast, if we "dice" the same Pt layer into several layers by inserting multiple ultra-thin Hf layers during the deposition process, the new Pt/Hf interfaces should introduce strong additional interfacial scattering of electrons and hence greatly enhance the averaged $\rho_{xx}$. The Pt crystal structure between the interfaces can be disrupted less than would be the case for uniform alloying with Hf [19], thereby better preserving the large intrinsic $\sigma_{SH}$ of Pt and better enhancing $\xi_{DL}^j$.

We sputter-deposited magnetic stacks of Ta 1.0/[Pt $d$/Hf 0.2]$_n$/Pt $d$/Co $t$/MgO 2.0/Ta 1.5 (numbers are layer thicknesses in nm) with $d$ = 0.4, 0.5, 0.6, 0.75, 1, 1.5, 2, and 4 nm, respectively. Here $n$ (≤ 7) is chosen to be the integer that can make the total Pt thickness closest to 4 nm under the constraint that the total Hf thickness is no more than 1.4 nm (note that the spin diffusion length $\lambda_s$ of the amorphous Hf is ~1 nm [22]). For the perpendicular magnetic anisotropy (PMA) samples, the Co thickness $t$ is 0.83 nm for $d \geq 1$ nm and 0.63 nm for $d \leq 0.75$ nm; for in-plane magnetic anisotropy (IMA) samples, $t$ is 1.3 nm for $d \geq 1$ nm and 0.93 nm for $d \leq 0.75$ nm. The samples were further patterned into 5×60 μm$^2$ Hall bars (see Fig. 2(a)) for resistivity and SOT measurements (see Supplemental materials [23]).

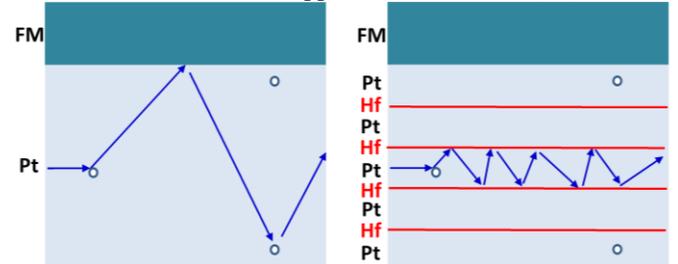

FIG 1. Schematic depiction of interfacial scattering enhancement of the resistivity.

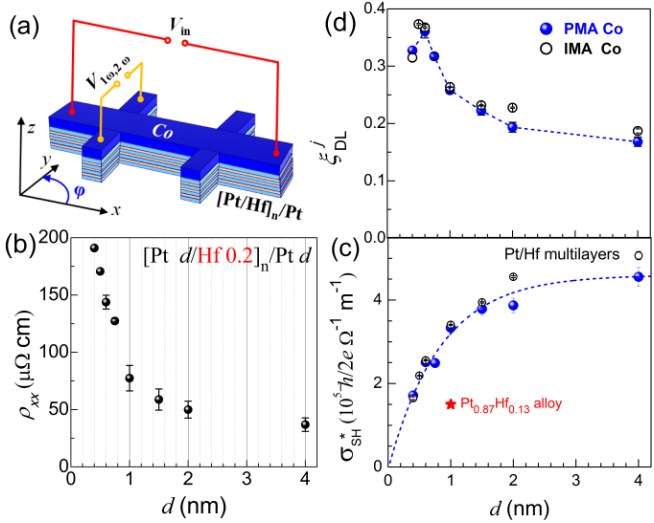

FIG 2. (a) Geometry and coordinates for the SOT measurements; (b) Measured resistivity of [Pt $d$/Hf 0.2]$_n$/Pt $d$ multilayers as a function of the "slice" thickness $d$ of the individual Pt layers; (c) The damping-like spin-torque efficiency ($\xi_{DL}^j$) and (d) the apparent spin Hall conductivity ($\sigma_{SH}^*$) determined from harmonic-response measurements for both PMA (blue dots) and IMA (black circles) samples plotted as a function of $d$. The dashed lines are guides to the eye. The red star denotes the value of $\sigma_{SH}^*$ for 4 nm of a spatially uniform Pt$_{0.87}$Hf$_{0.13}$ alloy [19].

As shown in Fig. 2(b), the average resistivity of the [Pt $d$/Hf 0.2]$_n$/Pt $d$ multilayer is increased from 37 μΩ cm for $d$ = 4 nm (pure Pt) to 191 μΩ cm for $d$ =0.4 ([Pt 0.4/Hf 0.2]$_7$/Pt 0.4). Compared to that achieved by alloying or impurity doping (~83 μΩ cm for Au$_{0.25}$Pt$_{0.75}$ and ~110 μΩ cm for Pt$_{0.85}$Hf$_{0.15}$)[18-21], this is a remarkable resistivity enhancement despite the fact that the 0.2 nm Hf insertions are too thin to be distinguishable by either x-ray diffraction/reflectivity or scanning tunneling electron microscopy (STEM)/electron dispersive spectroscopy (EDS) measurements (Supporting Information, Figs. S1-S3).

### 2.2 Magnifying spin torque by interfacial scattering

Figure 2(c) summarizes the values of $\xi_{DL}$ determined from harmonic response measurements [24,25] on both the PMA and IMA multilayers as a function of $d$, with good agreement between the two types of measurements (Supporting Information, Figs. S4 and S5). For both the PMA and IMA samples, $\xi_{DL}^j$ increases quickly from ~0.17±0.01 at $d$ = 4 nm (pure Pt) to a peak at $d$ = 0.6 nm and then drops slightly as $d$ increases further to 0.4 nm. The peak value of $\xi_{DL}^j$ = 0.37±0.01 for $d$ = 0.6 nm (i.e., [Pt 0.6/Hf 0.2]$_5$/Pt 0.6 multilayers) is significantly higher than the values reported for Pt$_{0.85}$Hf$_{0.15}$ ($\xi_{DL}^j$ ≈ 0.15)[19], Au$_{0.25}$Pt$_{0.75}$ ($\xi_{DL}^j$ ≈ 0.30)[18], β-W ($\xi_{DL}^j$ ≈ 0.2-0.3)[13,26] and β-Ta ($\xi_{DL}^j$ ≈ 0.12)[3]. We attribute the increase of $\xi_{DL}$ for Pt/Hf multilayers to the enhanced resistivity from interface scattering (see Fig. 2(b)). Based on the comparisons in Fig. S8 and Table S2, the giant $\xi_{DL}^j$ for Pt/Hf multilayers can provide very compelling current and energy efficiencies for spin torque applications, for instance for SOT-MRAMs, with a current efficiency superior to any other known material for practical applications.

The interesting peak behavior of $\xi_{DL}^j$ at $d$ ≈ 0.6 nm can be explained as due to a competition between $\rho_{xx}$ that increases quickly as a function of decreasing $d$ (Fig. 2(b)) and the apparent spin Hall conductivity, $\sigma_{SH}^* \equiv T_{int}\sigma_{SH} = (\hbar/2e) \xi_{DL}^j/\rho_{xx}$, that decreases sharply as $d$ decreases from 4 nm to 0.4 nm (Fig. 2(d)). This decrease in $\sigma_{SH}^*$ should be attributed partly to the enhanced attenuation of spin current in the Hf insertion layers and at the Pt/Hf interfaces. The amorphous Hf has a short $\lambda_s$ of ~1 nm and doesn't contribute to the generation of the spin current due to its negligble SHE [22]. Therefore, in the multilayers with small $d$ where the total Hf thickness reaches > 1 nm, there should be a strong attenuation of the spin currents that diffuse to the FM interface from the bottom Pt layers to exert a SOT. Each additional Pt/Hf interface could also contribute to spin backflow and spin memory loss that further reduce the SOTs [9,27-29]. In addition, the decrease of $\sigma_{SH}^*$ with $d$ could result in part from a strain-induced degradation of the Pt band structure (from a well ordered fcc texture to a nearly armorphous structure, see Fig. S1). Nevertheless, in the Pt/Hf multilayers $\sigma_{SH}^*$ is better preserved compared to that of uniformly doped Pt with Hf impurities. As shown in Fig. 2(d), $\sigma_{SH}^*$ for the 4 nm Pt$_{0.87}$Hf$_{0.13}$ is 1.5×10$^5$ ($\hbar/2e$)Ω$^{-1}$ m$^{-1}$ [19], which is a factor of 2 smaller than that of the Pt/Hf multilayers with similar Hf "concentration" (i.e. close to [Pt 1/Hf 0.2]$_3$/Pt 1). This suggests that such HM multilayers with strong interfacial scattering can be generally advantagous over the corresponding impurity doping because in the latter $\sigma_{SH}$ can be degraded more substantially by a stronger disturbance to the Pt band structure. We speculate that an enhancement of $\xi_{DL}^j$ beyond the value of 0.37 that we obtain here should be possible if the increase of resistivity, the insertion layer attenuation of spin current, and the insertion-induced Pt strain can be better balanced, for instance, by using an insertion material that has a longer $\lambda_s$, and an atomic radius closer to that of Pt (e.g., Ti) to minimize the disruption of the Pt crystal lattice and band structure.

### 2.3 Spin-torque switching of magnetization

Now we show that our optimal Pt/Hf multilayer with strong interfacial scattering, [Pt 0.6/Hf 0.2]$_5$/Pt 0.6, is a particularly compelling spin Hall material for SOT research and technological applications. As the first example, we show the switching of a PMA Co layer ($j_e$ = 1.7 × 10$^7$ A/cm$^2$, coercivity $H_c$ of 0.43 kOe) enabled by the giant $\xi_{DL}^j$ due to the SHE of the [Pt 0.6/Hf 0.2]$_5$/Pt 0.6 multilayer (Fig. S4). As an independent check of the effectiveness of the enhancement of $\xi_{DL}^j$ by Pt/Hf interfaces, we demsonstrate antidamping switching of in-plane magnetized SOT-MRAM devices with FeCoB-MgO MTJs. We fabricated two types of MRAM devices, Devices A and B. Each MRAM device consists of a 300 nm-wide spin Hall channel of [Pt 0.6/Hf 0.2]$_n$/Pt 0.6 ($n$ = 5 for Device A and 6 for Device B), an elliptical MTJ pillar of $Fe_{0.6}Co_{0.2}B_{0.2}$ 1.6/MgO 1.6/$Fe_{0.6}Co_{0.2}B_{0.2}$ 4 (190×45 nm$^2$ for Device A or 190×74 nm$^2$ for Device B), and protective capping layers of Pt 3/Ru 4 (see the schematic in Fig. 3(a) and cross-sectional STEM and EDS imaging results in Fig. S3). All devices were annealed at 240 $^o$C. For Device B, a 0.25 nm and a 0.1 nm Hf spacers were inserted at the bottom and top of the 1.6 nm FeCoB free layer, respectively, to suppress the magnetic damping constant ($\alpha$)[30] and reduce the effective magnetization ($4\pi M_{eff}$), thereby reducing the critical current for anti-damping switching [11]. The long axis of the elliptical MTJ pillars was along $y$ direction, transverse to the spin Hall channel and the write-current flow ($x$ direction). In Figs. 3(b)-3(f), we compare the magnetization switching behaviors, $\alpha$, and $4\pi M_{eff}$ of two representative MRAM devices without (Device A, red) and with (Device B, black) the two Hf spacers. Figure 3(b) shows the sharp switching minor loops of the MTJs under an in-plane magnetic field along the long axis of the MTJ pillar ($H_y$). The minor loops are artificially centered after subtraction of the dipole fields ($H_{offset}$ ≈150 Oe for Device A and 180 Oe for Device B) of the 4 nm $Fe_{0.6}Co_{0.2}B_{0.2}$ reference layers. $H_c$ of the free layer is 36 Oe for Device A and 9 Oe for device B. The apparent tunnel magnetoresistance ratio (~40% for Devices A and ~7% for Device B) is not very high, which is attributed to a large background resistance caused during the device fabrication process (i.e. the oxidization of the Ti adhesion layer between the MTJ pillars and the top Pt contact as indicated in Fig. S3).

Figure 3(c) shows the characteristic switching behavior of Devices A and B as the write current in the spin Hall channel is ramped quasi-statically (an in-plane field equal to $H_{offset}$ was applied along pillar long axis to compensate the dipole field from the reference layer). The MTJs show abrupt switching at write currents of 16 μA for Device A and 20 μA for Device B. Since thermal fluctuations assist the reversal of a nanoscale MTJ device during slow current ramps, we carried out ramp rate measurements (Fig. 3(d)). Within the macrospin model, the switching current $I_c$ should scale with the ramp rate ($\dot{I}$) following [31]

$$I_c = I_{c0}\left(1 + \frac{1}{\Delta}\ln\frac{\tau_0\Delta|\dot{I}|}{|I_{c0}|}\right) \quad (1)$$

Here $I_{c0}$ is the critical switching current in absence of thermal fluctuations, $\Delta$ the stability factor that represents the normalized magnetic energy barrier for reversal between the P and AP states, and $\tau_0$ the thermal attempt time which we assume to be 1 ns. By fitting to Eq. (1), we obtain $|I_{c0}| \approx$ 172±18 μA and $\Delta \approx$ 26 for Device A and $|I_{c0}| \approx$ 73±15 μA and $\Delta \approx$ 29 for Device B after averaging the critical currents for P→AP and AP→P switching. The small critical switching currents are consistently reproduced by other devices. Considering a parallel resistor approximation, the current shunted into the FeCoB free layer and Hf spacers ($\rho_{Pt/Hf}$≈144 μΩ cm, $\rho_{FeCoB}$≈$\rho_{Hf}$≈130 μΩ cm) can be estimated to be ≈ 0.2$I_{c0}$ for both devices (see Fig. S8 and Table S1). The critical switching density in the Pt spin Hall channel is therefore $j_{C0} \approx (1.0±0.1)\times10^7$ A/cm$^2$ for Device A (no Hf spacers) and $j_{C0} \approx (3.6±0.7)\times10^6$ A/cm$^2$ for Device B (with Hf spacers). Both the total critical switching and the low switching current density obtained from Device B are the lowest yet reported for any in-plane [2,10,11,20,26] or perpendicular [12] spin-torque MTJ (see Table 1).

According to the macrospin model, $j_{C0}$ for antidamping torque switching of an in-plane magnetized MTJ is given by $j_{C0} = (2e/\hbar)\mu_0 M_s t\alpha(H_c + 4\pi M_{eff}/2)/\xi_{DL}^j$ [32]. With $\alpha$ of 0.017 (0.011), $4\pi M_{eff}$ of 5.54 (1.94) kOe, and $M_s$ of 1240 emu/cm$^2$ for the magnetic free layer of Device A (B) as calibrated from ferromagnetic resonance (FMR) (see Figs. 3(e) and 3(f)) and vibrating sample magnetometry measurements on un-patterned thin film stacks, we estimate $\xi_{DL}^j$ to be ~0.29 for Device A and 0.17 for Device B.

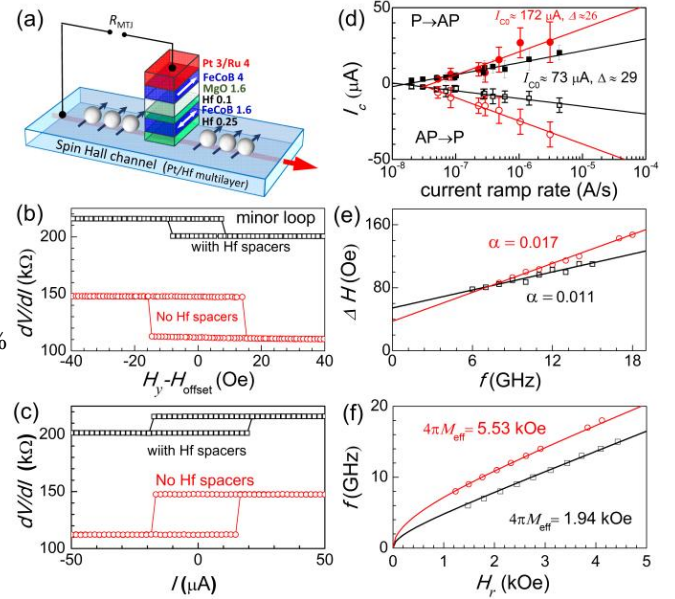

FIG 3. (a) Schematic of the 3-terminal MRAM device. (b) Minor loop for switching by an in-plane applied magnetic field, (c) Direct current switching loop, (d) critical current for P→AP (solid) and AP→P (open) switching as a function of current ramp rate, (e) FMR linewidth $\Delta H$ as a function of the resonance frequency $f$, (f) FMR resonance field $H_r$ for the 1.6 nm FeCoB magnetic free layers for Device A (red) and Device B (black). The solid lines in (d), (e), and (f) denote the best fits of data to Eq. (1), $\Delta H = \Delta H_0 + (2\pi/\gamma)\alpha f$, and $f = (\gamma/2\pi)\sqrt{H_r(H_r + 4\pi M_{eff})}$, respectively. $\Delta H_0$ and $\gamma$ are the inhomogeneous broadening of the FMR linewidth and the gyromagnetic ratio, respectively.

Table 1. Comparison of SOT-MRAM devices. Both the critical switching current ($I_c$) and the critical switching current density ($j_{c0}$) for our Pt/Hf multilayer device are the lowest among all spin-Hall materials demonstrated in room-temperature SOT-MRAM devices. Here [Pt/Hf]$_n$ represents the multilayers of [Pt 0.6/Hf 0.2]$_6$/Pt 0.6.

| | SOT device | $I_{c0}$ (mA) | $j_{c0}$ (MA/cm$^2$) | Refs |
|---|---|---|---|---|
| [Pt/Hf]$_n$ | In-plane MTJ | 0.073 | 3.6 | This work |
| W | In-plane MTJ | 0.15 | 5.4 | [11] |
| W | In-plane MTJ | 0.95 | 18 | [26] |
| Pt | In-plane MTJ | 0.67 | 40 | [10] |
| Ta | In-plane MTJ | 2.0 | 32 | [3] |
| Pt$_{0.85}$Hf$_{0.15}$ | In-plane MTJ | 0.56 | 14 | [20] |
| Ta | PMA MTJ | >20 | >50 | [12] |

The slight reduction of $\xi_{DL}^j$ for Device B compared to Device A is attributed to the spin current attenuation and possible reduction of the effective spin mixing conductance due to the insertion of the 0.25 nm Hf layer in Device B between the Pt/Hf multilayer and the FeCoB layer. Despite this reduction, this Hf spacer layer is still beneficial in that the suppression of $\alpha$ and the reduction of $4\pi M_{eff}$ for the free layer interface more than compensates for the decrease in $\xi_{DL}^j$. The value of $\xi_{DL}^j \sim 0.29$ for Device A is significantly higher than those previously obtained in similar studies for MRAM devices based on $\beta$-W ($\xi_{DL}^j$ = -0.15)[11], Pt$_{0.85}$Hf$_{0.15}$ ($\xi_{DL}^j$ = 0.098)[20], and Pt ($\xi_{DL}^j$ = 0.12)[30]. We do note that $\xi_{DL}^j$= 0.29 from the MRAM ramp rate experiment is ~20% less that the value determined from harmonic response measurement (see Fig. 2(c)). This difference may be partly attributed to an increased magnetic damping of nanoscale devices compared to thin film stacks due to, e.g., the ion-beam damage and the side-wall oxidation of the nanopillar during the device fabrication process. Tapering of free layer which was formed during the ion milling process due to the resist shielding effect (see more details in Fig. S3), can significantly increase the effective volume of the free layer of the MRAM device and lead to additional current shunting into the free layer. This current shunting into the tapering area has not been taken into account in our calculation. For the same reasons $\xi_{DL}^j$ of spin Hall materials is generally found to be underestimated in the ramp rate results of other nanoscale MRAM devices compared to in direct SOT measurements on micro-scale Hall bars [10,11,20] (e.g. for W, $\xi_{DL}^j$ is ~0.15 from MRAM ramp rate measurements and ~0.20 from bilayer spin-torque measurements [11]).

We point out the record-low critical switching current (current density) of the SOT-MRAMs based on Pt/Hf multilayers is a technologically interesting achievement. The 3-terminal SOT-MRAM is an advantageous current- and energy-efficient cache memory candidate because the separation of the read and write channels in the 3T geometry offers additional advantages over the conventional 2-terminal spin-transfer-torque geometry: e.g., unlimited endurance, faster write (sub-ns [11]), faster readout without read disturbance, lower write energy, and allowance for thick MgO barrier for enhanced TMR.

## Conclusion

In conclusion, we have demonstrated, from direct spin-torque measurements and also spin-torque switching experiments of magnetic layers with both perpendicular and in-plane magnetic anisotropy, that introducing additional interface electron scattering within Pt by inserting sub-monolayer layers of Hf can significantly increase $\xi_{DL}^j$. For example, we show an increase of $\xi_{DL}^j$ from ~0.17±0.01 for a simple 4 nm-thick single Pt layer to ~0.37±0.01 for a [Pt 0.6 /Hf 0.2]$_7$/Pt 0.6 multilayer despite the attenuation of spin current from Pt by the Hf insertion layers. Taking advantage of this interface-scattering-enhanced spin Hall ratio in the Pt/Hf multilayers, we demonstrate deterministic switching of IMA FeCoB-MRAM devices with a critical switching current of ~73 µA and critical switching current density of ~3.6×10$^6$ A/cm$^2$ in absence of thermal fluctuations, both of which are the lowest values yet known. Our optimized multilayer, [Pt 0.6/Hf 0.2]$_5$/Pt 0.6 (with $\xi_{DL}^j$ = 0.37, $\rho_{xx}$ = 144 µΩ cm), represents a highly-efficient generator of spin-orbit torque that is also compatible with integration technology (e.g., allowing easy growth with standard sputtering techniques on Si substrates) for development of low-power magnetic memories, oscillators, and logic. Our findings also provide a new strategy with the potential to magnify SOTs generated by other heavy metals, e.g., the low-resistivity Pd-Pt [17] or Au-Pt [18].

## Acknowledgements

This work was supported in part by the Office of Naval Research (N00014-15-1-2449), by the NSF MRSEC program (DMR-1719875) through the Cornell Center for Materials Research, and by the Office of the Director of National Intelligence (ODNI), Intelligence Advanced Research Projects Activity (IARPA), via contract W911NF-14-C0089. The views and conclusions contained herein are those of the authors and should not be interpreted as necessarily representing the official policies or endorsements, either expressed or implied, of the ODNI, IARPA, or the U.S. Government. The U.S. Government is authorized to reproduce and distribute reprints for Governmental purposes notwithstanding any copyright annotation thereon. This work was performed in part at the Cornell NanoScale Facility, an NNCI member supported by NSF Grant ECCS-1542081.